\documentclass[aps,prx,twocolumn,secnumarabic,nobalancelastpage,amsmath,amssymb,
nofootinbib]{revtex4-1}

\usepackage{graphics}      
\usepackage{graphicx}      
\usepackage{longtable}     
\usepackage{url}               
\usepackage{bm}  
\usepackage[explicit]{titlesec}
\titlespacing\section{0pt}{8pt}{4pt}
\usepackage{epstopdf}

\begin{document}
\title{Permutation Entropy and Statistical Complexity Analysis of Turbulence in Laboratory Plasmas and the Solar Wind}
\author{P.J. Weck, D.A. Schaffner, and M.R. Brown}
\affiliation{Swarthmore College, Swarthmore, PA 19081}
\author{R.T. Wicks}
\affiliation{NASA Goddard Space Flight Center, Greenbelt, MD 20771}
\begin{abstract}
The Bandt-Pompe permutation entropy and the Jensen-Shannon statistical complexity are used to analyze fluctuating time series of three different plasmas: the magnetohydrodynamic (MHD) turbulence in the plasma wind tunnel of the Swarthmore Spheromak Experiment (SSX), drift-wave turbulence of ion saturation current fluctuations in the edge of the Large Plasma Device (LAPD) and fully-developed turbulent magnetic fluctuations of the solar wind taken from the $\textit{Wind} $ spacecraft. The entropy and complexity values are presented as coordinates on the CH plane for comparison among the different plasma environments and other fluctuation models. The solar wind is found to have the highest permutation entropy and lowest statistical complexity of the three data sets analyzed. Both laboratory data sets have larger values of statistical complexity, suggesting these systems have fewer degrees of freedom in their fluctuations, with SSX magnetic fluctuations having slightly less complexity than the LAPD edge $I_{\rm sat}$. The CH plane coordinates are compared to the shape and distribution of a spectral decomposition of the waveforms. These results suggest that fully developed turbulence (solar wind) occupies the lower-right region of the CH plane, and that other plasma systems considered to be turbulent have less permutation entropy and more statistical complexity. This paper presents the first use of this statistical analysis tool on solar wind plasma, as well as on an MHD turbulent experimental plasma.
\end{abstract}
\maketitle

\section{Introduction}
Since Bandt and Pompe introduced their probability distribution based on ordinal patterns in arbitrary time series in 2002 \cite{bandt2002}, their methodology has found a wide variety of applications, from tracking the effects of anesthetic drugs on the brain \cite{li2010,jordan2008,olofsen2008} to informing economic policy \cite{bariviera2013,zunino2010,zunino2011} to various other areas \cite{kowalski2007,soriano2011,saco2010,suyal2012,sun2010}. In 2007, Rosso \textit{et al} applied the ordinal pattern distribution of Bandt and Pompe to a time series analysis using the complexity-entropy plane, or ``CH plane'', capable of differentiating between periodic, chaotic, and stochastic systems \cite{rosso2007}. The CH plane has been used to determine the statistical character of fluctuations in several plasma systems, including magnetic flux ropes \cite{gekelman2014} and electron heat transport \cite{maggs2013}. However this approach has yet to be extended to the study of dynamical MHD turbulence, either in the solar wind or in laboratory MHD plasma. The purpose of this paper is to provide the CH plane coordinates for these turbulent systems and compare to previous results, as well as to further the interpretation of this analysis tool for the study of turbulent plasma systems.

We compute the values of permutation entropy and Jenson-Shannon complexity for magnetic fluctuations of SSX and the solar wind and the ion saturation current ($I_{\rm sat}$) fluctuations of the LAPD edge. These values are then used as coordinates for placement in the CH plane for comparison among each other as well as to known chaotic and stochastic models. The results show that the magnetic solar wind fluctuations have the highest level of permutation entropy and lowest level of complexity, occupying a position on the lower right region of the CH plane, nearest that of pure white noise, which has zero complexity and maximal entropy. This result suggests that fully developed turbulence, as the solar wind is thought to represent, can be identified by its proximity to maximal stochasticity on the CH plane. The LAPD edge fluctuations have the highest level of complexity of the three measured data sets and occupies the middle region in permutation entropy. Previous work has shown that the LAPD drift-wave turbulence may be dominated by non-linear interactions of relatively small numbers of modes, and thus tend to exhibit more chaotic, complex behavior \cite{pace2008}; thus, its coordinates occupy a position closest to known chaotic maps. Finally, the SSX fluctuations exhibit a level of complexity in between the other two plasmas. This suggests that the SSX plasma has more degrees of freedom in its fluctuations than the LAPD drift-wave plasma, but is not fully-developed turbulence or is constrained by the laboratory boundaries. The permutation entropy of the SSX magnetic fluctuations is relatively high or low depending on whether fluctuations in dB/dt ($\dot{B}(t)$) or temporally integrated B-field fluctuations (B(t)) are analyzed. This difference suggests that the level of entropy of a time series may be related to the rate of decrease in power as frequency increases.

It should be emphasized that the goal of this comparative study at this stage is to highlight the variations in outcomes of using this particular analysis tool, rather than attempting to unravel differences in the physical mechanisms underlying each dataset. In a sense, the work presented here was designed to be as physics-blind as possible. However, through study of how various mechanisms manifest in the complexity-entropy plane, a comprehensive physical understanding of each system can be pursued.

The MHD wind tunnel configuration of the Swarthmore Spheromak Experiment (SSX) consists of a plasma gun which injects a spheromak of magnetized plasma into an $\sim1$ meter long cylindrical copper flux conserver~\cite{gray2013}. Probes embedded in the chamber collect data on turbulent fluctuations in $\dot{B}$ as the plasma evolves down the length of the tube, eventually relaxing into a Taylor state~\cite{gray2013,schaffner2014a,schaffner2014b,schaffner2014c}. After injection the plasma is completely dynamical, as there is no guide or vacuum field in the body of the chamber. The $\dot{B}$ fluctuation signals for SSX  were recorded by a 16-channel, 3-direction, single-loop pickup coil probe array embedded in the midplane of the cylindrical wind tunnel, with a $65$ MHz sampling rate and 14 bit dynamic range. By varying the amount of magnetic flux through the core of the gun, referred to here as ``stuffing flux'', the magnetic helicity of the injected plasma can be finely controlled~\cite{schaffner2014b}. Magnetic helicity corresponds to the degree of twistedness in the magnetic field, so varying injected helicity affects the resulting turbulent dynamics of the plasma as it evolves towards a relaxed Taylor state. 

We compare observations from the {\it Wind} spacecraft in the turbulent solar wind to the laboratory plasma experiments. The {\it Wind} spacecraft provides high-cadence magnetic field observations of the solar wind using the MFI~\cite{lepping1995} from the L1 Lagrangian point between the Earth and the Sun. Measurements are made 11 times per second using a flux gate magnetometer and then averaged to 3s to remove the spacecraft spin signal from the data. Flux gate measurements provide a DC magnetic field observation by measuring the bias required for no current to flow in a coil of wire while subject to a changing magnetic field. Thus the observations are equivalent to the B(t) observations made in SSX (but not $\dot{B}$). The solar wind is highly variable but there are broadly two types of solar wind: fast wind (V $>$ 600 km/s) which is emitted from open coronal field lines and is typically low density ($< 5$ protons/cm$^3$), has few large scale structures and has high amplitude but less developed turbulence, and slow wind, ($V < 500 $km/s) which is typically found in the ecliptic plane and originates from more complex coronal magnetic topology and is denser and more structured than the fast wind with more evolved but lower amplitude turbulence~\cite{tu1990,bruno2013}. Here we use multi-day long intervals of a fast wind stream (Jan 14 - Jan 21 2008) and a slow wind stream (Jan 24 - Jan 29 2010) with large scale magnetic fluctuations on the order of 10 nT. 

Edge $I_{\rm sat}$ fluctuations on the LAPD~\cite{gekelman1991} were taken using biased Langmuir probes inserted radially from the cylindrical edge of the plasma device. Signals were sampled at 1.5MHz from a radial location of 26cm~\cite{schaffner2012}. The fluctuations in the edge are shown to be dominated by drift-wave modes due to the pressure gradient that develops between the plasma core and the chamber wall~\cite{maggs1996}.

\section{Permutation Entropy and the CH Plane}
The permutation entropy of an arbitrary time series is defined in terms of a window length called the embedding dimension $n$. The embedding dimension determines the size of patterns investigated in calculating the entropy and complexity of the series. The instances of each ordinal patterns of that size are counted in order to associate an ordinal pattern probability distribution with the time series, from which the calculation of entropy and complexity is straightforward.

For embedding dimension $n$, the probability distribution introduced by Bandt and Pompe consists of the frequencies of occurrence of all possible length $n$ ordinal patterns in segments of $n$ consecutive terms from an arbitrary time series \cite{bandt2002}. In their methodology, a length $n$ ordinal pattern is defined for a segment $s = ( x_t,x_{t+1},\ldots,x_{t+(n-1)} )$ of the time series as the permutation $\pi$ of the index set $\{0,1,\ldots,n-1 \}$ corresponding to the ranking of the $x_i$ in ascending order, namely $x_{\pi_t}< x_{\pi_{t+1}}<\ldots< x_{\pi_{t+(n-1)}}$. In order to guarantee a unique result, if $x_i = x_j$ where $i<j$, then in the ranking $x_i <x_j$. For example, if $x_0 = 5$,  $x_1= -2$,  and $x_2 = 0.33$ are three consecutive terms in the time series, then since $x_1 < x_2 < x_0$, the ordinal pattern for this segment is the permutation $\pi = (1,2,0)$.  Given a time series of length $L$, the corresponding ordinal pattern probability distribution $P= \{p(\pi) \}$ is defined in terms of all $L-n+1$ length $n$ segments $s$ in the series and all $n!$ permutations $\pi$ of order $n$ by
\begin{align}
p(\pi) = \frac{|\{s: \text{$s$ has ordinal pattern $\pi$}\}| }{L-n+1}. 
\end{align}
where $|\ldots|$ denotes cardinality. The permutation entropy $PE$ is defined as Shannon's information entropy for this ordinal pattern probability distribution, or
\begin{align}
PE  = -\sum^{n!} p(\pi) \log p(\pi)
\end{align}
where the $\log$ is base two.

Instead of considering consecutive points in calculating the ordinal pattern probability distribution for a time series, an embedding delay $\tau$ can be used to sample ordinal patterns on a larger time scale, thereby placing a lower limit on the temporal size of structures resolved, consequently limiting the maximum associated frequency. Embedding delays can be implemented as a simple sub-sampling of data in which only $L/\tau$ values of the time series are considered~\cite{maggs2013,gekelman2014} or all portions of the original time series can be used~\cite{bandt2005}, a method referred to here as the length-preserving method. For example, for an embedding delay $\tau=10$ using the former approach, a new time series $X'$ of length $L'=\frac{1}{10}L$ is generated by selecting every tenth value of the original series $X$ and the ordinal pattern probability distribution calculated for that series in the usual manner. In the length-preserving method, segments $ ( x_t,x_{t+10},\ldots,x_{t+10(n-1)} ) $ of $X$ are used to calculate the ordinal pattern probability distribution, where $t$ runs from $1$ to $L-10(n-1)$, thereby including the $9/10$ths of the dataset thrown out in the first method. Which method is used depends in part on the length of the record in question. Unless  $L' \gg n!$, the first method may not yield reliable statistics~\cite{gekelman2014}, and the length-preserving method thus appears preferable.

While the permutation entropy quantifies the randomness in an arbitrary time series, a measure of statistical complexity such as the Jensen-Shannon complexity is required to quantify the degree of correlational structure in the time series~\cite{rosso2007}. The Jensen-Shannon complexity, or $C_{JS}$, is a functional of the discrete distribution $P$ of $N$ probabilities associated with the time series. Once normalized such that $0 \leq C_{JS} \leq 1$, 
\begin{align}
C_{js}[P] = -2\frac{S \left[ \frac{P+P_e}{2} \right] - \frac{1}{2}S[P]-\frac{1}{2}S[P_e] }{\frac{N+1}{N} \log(N+1)-2 \log(2N)+\log(N)}H[P]
\end{align}
Where $S$ is the Shannon entropy, $H$ is the normalized Shannon entropy $S/S_{\text{max}}$, and $P_e=\{\frac{1}{N}, \ldots, \frac{1}{N} \}$ is the uniform distribution. When analyzing time series using the CH plane methodology of Rosso \textit{et al}, this measure of statistical complexity is evaluated by associating with the time series the ordinal pattern probability distribution of Bandt and Pompe, so that $N=n!$, $S[P]=PE$ and $H[P]=PE_{\text{norm}}=PE/\log(n!)$. The statistical nature of time series is then evaluated by comparing their positions on the CH plane, $PE_{\text{norm}} \times C_{JS}$.

\begin{figure}[!htbp]
\centerline{
\includegraphics[width=8.5cm]{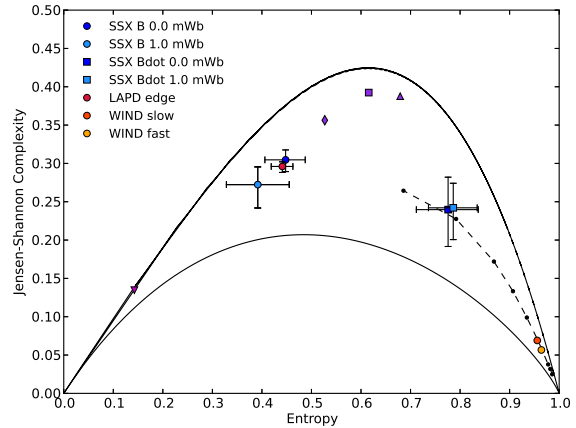}}
\caption{\label{fig:CHplane} The $n=5$ CH plane  with SSX $\dot{B}$ and $B$ (time-integrated from $\dot{B}$) data for two injected helicities, \textit{Wind} fast and slow stream $B$ data, LAPD edge plasma ion saturation current signals, and paradigmatic chaotic, periodic, and stochastic systems for comparison. The diamond, square, and triangular purple markers represent chaotic skew tent, Henon, and logisitic maps, respectively. The downward pointing triangle marks the position of the Sine function, and stochastic fBm points are shown in black. Crescent shaped curves show the maximum and minimum possible $C_{JS}$ for a given $PE_{\text{norm}}$. Error bars indicate standard deviation from the ensemble average. Solar wind bars are smaller than the displayed size of the markers.}
\end{figure}

For interpretation the CH plane coordinates of experimental data, it is useful to compare them to well-known stochastic and chaotic models. To illustrate the regions of the plane corresponding to paradigmatically stochastic and chaotic dynamics, stochastic fractional Brownian motion (fBm) and chaotic Henon, skew tent, and logistic maps are included in the $n=5$ CH plane shown in Figure~\ref{fig:CHplane} \cite{rosso2007}. The range of fBm points displayed was generated by varying the corresponding Hurst exponent between 0 and 1, thereby scanning the degree of correlation between successive increments of motion from strong negative correlations to positive correlations. A time series generated by a sine function is included as well, illustrating the low entropy, low complexity domain occupied by simple mathematical functions. Note that pure white noise occupies the $PE_{\text{norm}}=1$, $C_{JS}=0$ corner of the plane. Also shown in Figure~\ref{fig:CHplane} are curves demarcating the minimum and maximum complexity bounds of the CH plane. The nature of the dependence of $C_{JS}$ on $H$ constrains the possible values of the former as a function of the latter to fall between these curves~\cite{lopez1995,calbet2001}.
\begin{figure}[!htbp]
\centerline{
\includegraphics[width=8.5cm]{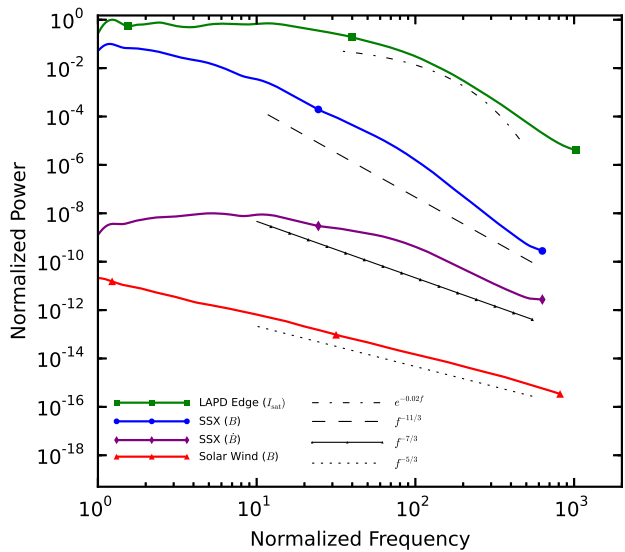}}
\caption{\label{fig:spectra} Spectra of LAPD edge $I_{\rm sat}$ fluctuations, SSX magnetic fluctuations and solar wind magnetic fluctuations. Each spectrum is normalized to a different time scale: LAPD is normalized to 500Hz; SSX is normalized to 12.8kHz; solar wind is normalized to 165$\mu$Hz. Analytic forms for an exponential, a -11/3 power-law, and a -5/3 power-law (Kolmogorov) are indicated by the black lines. The power scale is arbitrary as the emphasis here is on the shape of the curves, not the relative power content of the spectra. The three spectra indicate a clear transition from exponential-like to power-law like broadband spectra.}
\end{figure}
\section{CH comparison of SSX, WIND and LAPD data}
SSX magnetic fluctuations were analyzed over a 20 $\mu$s window during the stationary period of the discharge corresponding to 1,300-sample records and averaged over 40 shots. Actual magnetic field fluctuations, $B$, are obtained by integrating the $dB/dt$ signal over time. The normalized permutation entropy and Jensen-Shannon complexity were calculated for each series, using $n=5$ in order to satisfy the common condition $L > 5n!$, as recommended in \cite{amigo2008} and \cite{riedl2013}. The length-preserving embedding delay method was employed to preserve this condition after sub-sampling. An embedding delay of $\tau=8$ was used to filter frequencies above 9MHz to avoid contamination from a high frequency noise mode, but small enough compared to the record length to avoid artificial numerical effects we found to be associated with small L to $\tau$ ratios. The average position of series from all three directions of the inner $4$ probe coils at each of two helicity settings is shown in blue in Figure~\ref{fig:CHplane}. Error bars indicate standard deviations from the 40-shot ensemble average.  

Figure~\ref{fig:CHplane} also shows the positions of both fast and slow stream magnetic fluctuations in the solar wind.  The fast stream magnetic signal from \textit{Wind} consisted of almost $230,000$ values, and the slow stream signal of over $170,000$. Since both signals were highly stationary, a set of subseries could be treated as an ensemble. The length of subseries $L_{wind}$ was chosen in conjunction with the embedding delay $\tau_{wind}$ so as to satisfy the condition $L_{wind}/\tau_{wind} = L_{ssx}/\tau_{ssx}$. Entropies and complexities were averaged over $20$ subseries each $11,375$ values in length for the fast stream signal and $15$ subseries of $11,375$ values for the slow stream. Delays of $\tau_{wind} = 70$ were used, which limits the  upper frequency range of the dynamics under investigation to well within the inertial range. Error bars are within the range of the marker.

Previous work using frequency spectra has suggested that the edge fluctuations of magnetized plasmas in the LAPD and other devices are chaotic in nature \cite{maggs2012}. The CH coordinate of the LAPD edge plasma shown in Figure 1 in red was averaged over $25$ shots and $5$ sections of $1000$ values for each shot with no embedding delay. 

The relative coordinates of each measurement show that the solar wind magnetic fluctuations at $1$ AU are the most stochastic-like of the three with permutation entropy and complexity values of (H=0.964, C=0.057) for fast and (H=0.956, C=0.069) for slow wind, both close to that of pure white noise and more random than even classical Brownian motion, or fBm with Hurst exponent of $1/2$ (fBm models have also been explored as a potential model for turbulent fluctuations in the solar wind and the magnetosphere~\cite{watkins2005}). The fast stream signal exhibits slightly more entropy and less complexity than the slow stream signal; it is as yet unclear whether this slight difference has a physical meaning, however. Although it has been well documented that the solar wind exhibits well-developed turbulence~\cite{bruno2013}, this is the first time that developed MHD turbulence in an astrophysical plasma has been identified based on this complexity-entropy plane analysis or compared in this manner to other plasma sources.

Conversely LAPD edge fluctuations are the most chaotic-like with coordinates of (H=0.441, C=0.296), closest of the three measurements to the chaotic models at the top of the CH plane. Although the complexity values for the full LAPD edge are slightly less than that observed in smaller drift-wave experimental setups~\cite{maggs2013}, the relatively high complexity compared to the other measurements suggests a larger contribution from chaotic dynamics, likely associated with the non-linear interaction of the drift-wave modes~\cite{maggs2012}. 

Finally, SSX magnetic fluctuations have entropy/complexity values of (H=0.776, 0.786; C=0.24, 0.242 ) for $\dot{B}(t)$ data (0.0 and 1.0 mWb stuffing fluxes) and (H=0.448, 0.392; C=0.305, 0.272) for $B(t)$ (same stuffing fluxes). The complexity values are in between that of LAPD and the solar wind, while the permutation entropy values differ substantially whether $dB/dt$ or $B$ is used. Naturally, this suggests that the magnetic fluctuations have a slightly more stochastic character than the density fluctuations of the LAPD edge, but do not reach the level of stochasticity of solar wind fluctuations. The large gap in entropy may be associated with the nature of the power spectrum as will be discussed next.

The results of the CH plane analysis can be compared to a typical power spectrum analysis. Figure~\ref{fig:spectra} shows the wavelet-generated power spectra~\cite{torrence1998} for the time series under investigation. Each spectrum is normalized to its minimum frequency in order for each curve to be placed on the same axis; this allows for the overall shape of the spectra to be directly compared. Furthermore, each curve is placed arbitrarily on the y-axis. Each spectrum is also cut-off at the frequency associated with the embedding delay used in the CH plane analysis. The LAPD spectrum shows the most exponential-like ($\sim e^{\tau f}$) shape while the solar wind spectrum is the most power-law like ($\sim f^{-\alpha}$). SSX $\dot{B}$ and $B$ spectra are in between and have slightly more power-law behavior. Since exponential spectra are typically identified with low-dimensional chaotic behavior~\cite{maggs2012}, the range in spectra mirror the results of the complexity analysis. The most exponential spectrum (LAPD) has the highest level of complexity while the most power-law like (solar wind) has the least complexity. The spectra also shed light on interpretation of the permutation entropy. The steepest spectra in Fig~\ref{fig:spectra} is the SSX $B$ spectrum; the corresponding time series also has the lowest amount of entropy. The LAPD data, if it were compared to a power-law slope, would have the second steepest spectrum while the SSX $\dot{B}$ spectrum is third, and finally the solar wind is the shallowest. This ordering is consistent with the relative magnitudes of permutation entropy for each time series. These results suggest that the permutation entropy is associated with the overall distribution of frequency power content of the time series, while the exponential versus power-law shape is associated with the level of complexity. It is clear that though each of these spectra is considered broadband and would perhaps be described as turbulent, the CH plane analysis reflects the different physical mechanisms which produce the fluctuations.

Next, the meaning of turbulence in the context of the nature of these fluctuations can be explored. The coordinates of solar wind magnetic fluctuations on the CH plane would suggest that fully developed turbulence should occupy a region close to the most stochastic limit. Meanwhile, fluctuations in a laboratory setting, while often referred to as turbulent (drift-wave turbulence for LAPD, MHD turbulence for SSX) may not be truly turbulent, or considered only weakly turbulent. Instead there appears to be a limit on how turbulent these fluctuations can be whether it is due to a limit on the number of modes associated with the fluctuations (as is thought to be the case in the LAPD~\cite{maggs2013}) or whether there is a limit on how much power can be distributed to higher frequencies (or smaller scales). In SSX, this latter issue may arise due to boundary or temporal development limits, both of which are not encountered by solar wind plasma (but may be relevant for the more bounded turbulent system of the magnetosheath~\cite{SahraouiPRL2006,YordanovaPRL2008}, for example). The results of the CH plane analysis highlight that more work is needed to push laboratory plasma turbulence research into the fully developed regime.
 
Finally, some discussion of how this analysis may be related to the typical measure of degrees of freedom in a turbulent plasma---Reynolds number---is warranted. Reynolds number, whether in reference to flow or magnetic turbulence (i.e. $Re$ or $R_{m}$), can be defined as the ratio of energy injection scale to energy dissipation scale in a turbulent cascade, and as such, can be interpreted as the number of degrees of freedom available to the system (or in other words, how many different scales energy can occupy between input and dissipation). The magnetic Reynolds number for the solar wind is typically on the order of $1 \times 10^7$ while SSX magnetic Reynolds numbers have been calculated (based on typical length scales and assuming Spizter resistivity as the dissipative mechanism) to be on the order of $1 \times 10^2$. Thus, Reynolds number shows a separation between solar wind data and SSX data though only in one dimension and qualitatively matches the difference in degrees of freedom suggested by the CH analysis. A complication arises when the LAPD data is introduced for comparison. Reynolds numbers are predicated on the separation of energy injection and dissipation scales. However, drift-wave turbulence may not have a clear separation of scales as energy can potentially be injected or dissipated at different scales~\cite{friedman2012}, and thus a Reynolds number may have less meaning in this case. The complexity-entropy analysis performed here, on the other hand, does not rely on any specific physics model and thus can be used to compare disparate systems.

\section{Conclusion}
In this paper, spectrally-broadband magnetic fluctuations in laboratory and astrophysical plasmas have been analyzed for the first time using the ordinal pattern-based CH plane introduced by Rosso \textit{et al}. Comparing the relative coordinates of drift-wave, MHD wind tunnel, and solar wind plasmas, it was found that the three systems occupy different regions of the CH plane, suggesting that despite the broad-band spectra exhibited by all these systems, the CH analysis is capable of highlighting differences in the underlying nature of the fluctuations, particularly among drift-wave, partially developed, and fully developed turbulence. Drift-wave turbulence is thought to be a result of the nonlinear interactions of relatively few modes while fully developed turbulence contains too many modes to distinguish; it appears that the entropy-complexity analysis of these magnetized plasmas effectively highlights the number of degrees of freedom of the system in question. In particular, the smaller number of modes generating drift-wave turbulence in LAPD edge plasmas are revealed by the low-middle entropy and middle-range complexity of that system, while the high entropy and low complexity of magnetic fluctuations in the solar wind may reflect the multitude of degrees of freedom active in that system. The analysis also showed that variations in permutation entropy maybe be related to power-law scaling of the spectra; in other words, permutation entropy may be proportional to the eveness of energy distribution among spectral frequencies. Based on the relative CH positions of SSX MHD wind tunnel and $\textit{Wind}$ data, although SSX is on its way towards the highly stochastic turbulence in the solar wind, this analysis indicates that further steps are needed for SSX  to more accurately model solar wind turbulence. The confined nature of the experiment and short lifetimes involved are both potential contributors to the discrepancy in CH positions. Other than the boundary conditions imposed by astrophysical bodies, the solar wind is an unconfined and extremely long lived plasma. Whether one or both of these parameters could be varied to reduce the complexity and increase the entropy of SSX to that of the solar wind is an open question. In any case, the CH methodology has provided us with another avenue for comparing and understanding turbulence in plasmas.

The authors gratefully acknowledge useful discussions with Jim Maggs, George Morales, Walter Gekelman, Brett Friedman, Troy Carter and Danny Guice. This work is supported by DOE OFES and NSF CMSO.

\end{document}